%% file: aa.tex
\documentclass[structabstract]{aa}  
\usepackage{graphicx}
\usepackage{txfonts}
\usepackage{natbib}
\begin{document}
\titlerunning{Warm Gas at 50~AU around Herbig Be star HD~100546}
\authorrunning{Goto et al.} 

 \title{Warm gas at 50 AU in the disk around Herbig Be star HD~100546
\fnmsep\thanks{Based on data collected in the course of CRIRES
                 program [084.C-0605] at the VLT on Cerro
                 Paranal (Chile), which is operated by the
                 European Southern Observatory (ESO).}
}

   \author{M. Goto\inst{1} \and
G. van der Plas\inst{2} \and
M. van den Ancker\inst{3} \and
C. P. Dullemond\inst{1,4} \and
A. Carmona\inst{5} \and
Th. Henning\inst{1} \and
G. Meeus\inst{6} \and
H. Linz\inst{1} \and
B. Stecklum\inst{7}}

\institute{Max-Planck-Institut f\"ur Astronomie,
  K\"onigstuhl 17, D-69117 Heidelberg, Germany
\email{mgoto@mpia.de}
\and
                 Astronomical Institute ``Anton Pannekoek'' at
  the University van Amsterdam, The Netherlands
\and
                 ESO, Karl-Schwarzschild-Stra\ss e 2,
                 D-85748 Garching bei M\"unchen, Germany
\and
Institut f\"ur Theoretische Astrophysik, Universit\"at
Heidelberg, Albert-Ueberle-Str. 2, D-69120 Heidelberg, Germany
\and
                 ISDC Data Centre for Astrophysics \& Geneva
  Observatory, University of Geneva, chemin d'Ecogia 16, 1290
  Versoix, Switzerland
\and
                  Universidad Aut\'onoma de Madrid,
                  Cantoblanco, 28049 Madrid, Spain
\and
                 Th\"uringer Landessternwarte Tautenburg, Sternwarte 5,
                 D-07778 Tautenburg, Germany
}

   \date{}
 
   \abstract 
{ The disk atmosphere is one of the fundamental elements of
  theoretical models of a protoplanetary disk. However, the
     direct observation of the warm gas ($\gg$100~K) at large
     radius of a disk ($\gg$10~AU) is challenging, because the line
     emission from warm gas in a disk is usually dominated by
     the emission from an inner disk. }
   {Our goal is to detect the warm gas in the disk atmosphere
     well beyond 10~AU from a central star in a nearby disk
     system of the Herbig~Be star HD~100546.}
   {We measured the excitation temperature of the vibrational
    transition of CO at incremental radii of the disk from the
    central star up to 50~AU, using an adaptive optics system
    combined with the high-resolution infrared spectrograph
    CRIRES at the VLT. }
{  The observation successfully resolved the line emission with
    0\farcs1 angular resolution, which is 10~AU at the distance
    of HD~100546. Population diagrams were constructed at each
    location of the disk, and compared with the models calculated 
taking into account the optical depth effect in LTE condition. 
The excitation temperature of CO is
    400--500~K or higher at 50~AU away from the star, where the
     blackbody temperature in equilibrium with the stellar
      radiation drops as low as 90~K. This is unambiguous
    evidence of a warm disk atmosphere far away from the central
    star.}
{} 

   \maketitle
%
\section{Introduction}

 A warm disk atmosphere is a natural consequence of a disk
  externally heated by stellar radiation
  \citep[e.g.][]{cal91,aik02}. Together with a somewhat cooler mid-plane
  and a disk inner-rim, it is one of the fundamental elements of
  the current theoretical models on which our understanding of
  a protoplanetary disk relies \citep[e.g.][]{dul01}. The
  emission from warm gas ($\gg$100~K) in a protoplanetary disk
  has been best observed in the vibrational transitions of CO
  \citep[e.g.][]{got06,bri07,naj07} and other molecules
  \citep[e.g.][]{wei00,car04}. It is hard to estimate the
  physical properties of the disk atmosphere from this
    spatially unresolved line emission alone, however, because the
  emission from the warm gas is dominated by the strong
  radiation from the hot inner disk closer to the central star
  \citep{bri07, pla09}. Sub-mm spectroscopy does probe the
  molecules in the outer disk ($>$100~AU) through the rotational
  transitions, but mostly  cool gas of the temperature
    15--30~K \citep[e.g.;][]{den05,pan10}.  In order to measure
  the hot gas in the disk atmosphere at radii of 10 to 100~AU,
  one requires (1) high angular resolution to exclusively
  observe the faint emission from the outer disk without being
  affected by the strong emission from the inner disk, (2) a
  proper spectroscopic probe that is sensitive to the warm disk
  atmosphere, many of which are in the infrared regime, and (3)
  the multiple transitions of that spectroscopic probe to
  measure the gas temperature quantitatively.

 \input{./t1.tex}

 The goal of our observation is to measure the rotational
  excitation temperature of CO fundamental band at 50~AU away
  from a star, using an adaptive optics system to isolate the
  emission of the outer disk from that of the inner rim. The
  nearest Herbig~Be star to the solar system, HD~100546
  \citep[B9.5Vne;][]{hou75}, is used as a testbed, because of
  its close distance \citep[$d=$103~pc; ][]{anc98} and high
  luminosity \citep[$L_\ast=26L_\odot$;][]{tat11}. HD~100546 is
  a 10~Myr-old \citep{anc98}, flared disk system \citep{meu01}
  \citep[$M_{\rm disk}>10^{-3}M_\odot$; ][]{pan09} inclined away from
  the observers by 40\degr--50\degr~\citep[e.g.,][]{ard07}, with
  two spiral arms seen at 200~AU in the scattered light
  \citep{gra01,gra05}.
The dust disk is truncated at 10~AU and inward
  \citep{bou03,ack06}. The gas and the dust at the inner rim are
  warm, 200~K and 1000~K, respectively, under the direct
  irradiation of the central star \citep{bou03,bri09}. The
  vibrational transition of CO and H$_2$ have been spatially
  resolved \citep{bri09,pla09,car11}, but the disk atmosphere
has not been studied in depth, because the emission from the outer
  disk was not properly separated from that of the inner rim. In
  this paper, we will follow up these studies, and
  measure the gas temperature in the disk atmosphere
  quantitatively.

\section{Observations}

The observation was carried out with CRIRES \citep{kau04} at the
VLT on 29 and 30 April in 2010 in the open-time program
084.C-0605. The adaptive optics system MACAO \citep{bon04} was
used to feed nearly diffraction-limited images to the
spectrograph using HD~100546 as a wavefront reference.  The slit
width was 0\farcs2, providing spectra with a velocity resolution
of 3~ km~s$^{-1}$ ($R$=100,000). The spectroscopy covers the
wavelength from 4.588~$\mu$m to 5.004~$\mu$m [$R$(16)--$R$(0),
$P$(1)--$P$(27)] continuously with six grating settings. The
spectra were obtained with the observing template {\it
  CRIRES\_spec\_obs\_SpectroAstrometry}. The slit was oriented eight
position angles to increase the spatial coverage of the disk. 
We here used the subset of the data with the slit
aligned to the major axis of the disk projected onto the sky
\citep[P.A.=145\degr;][]{ard07}, and anti-parallel to it
(P.A.=325\degr) to test the consistency of the detected line
emissions, and the measurement of the excitation temperature of
CO. The telescope was nodded by 10\arcsec~ every two exposures
to subtract the thermal emission from the sky. The total
integration time was 4 minutes for one grating setting. A
spectroscopic standard star HR~6556 (A5\,III) was observed at
close airmass with HD~100546 with the same instrument settings.

\begin{figure*}
\includegraphics[width=0.38\textheight,angle=0]{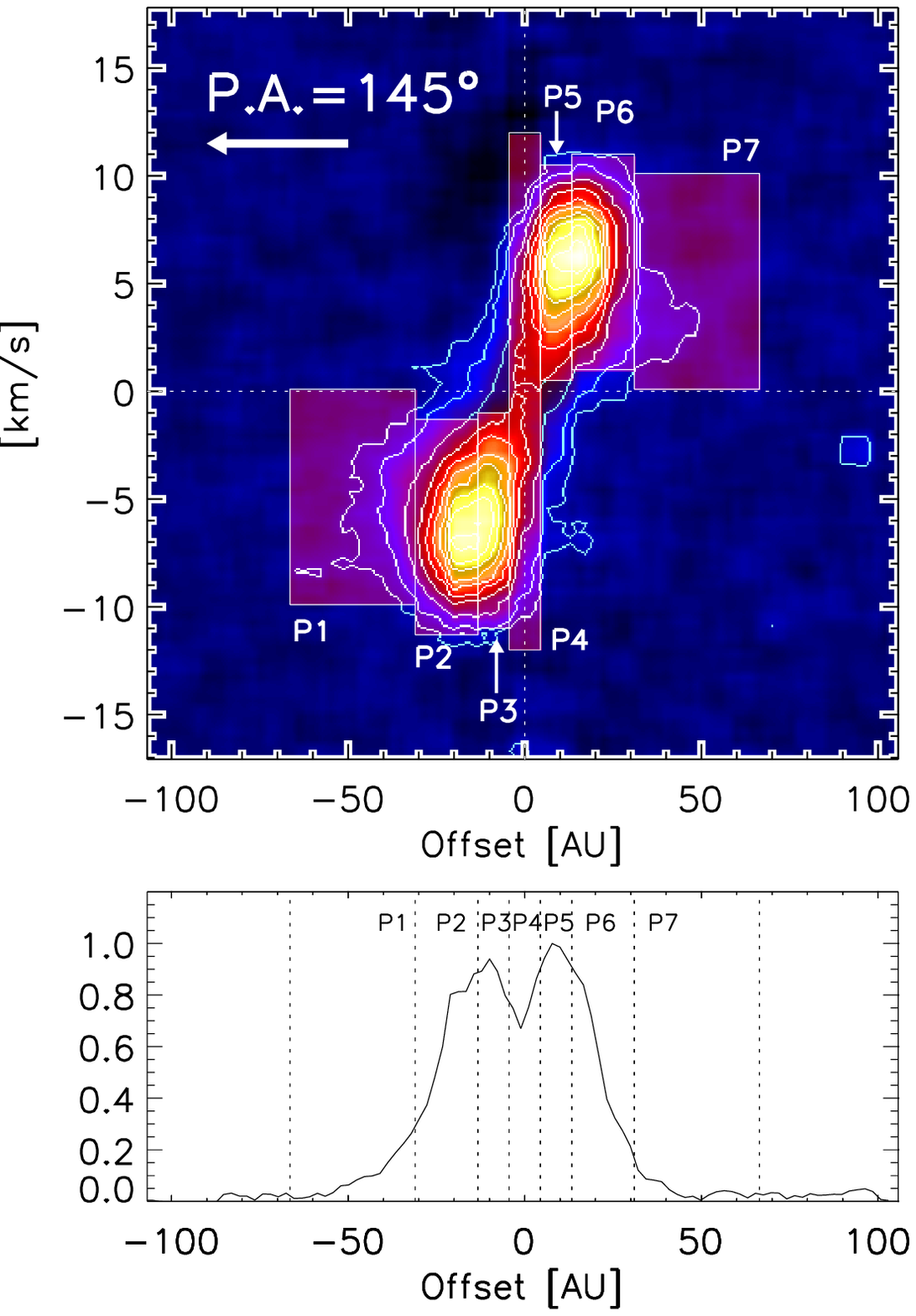} 
\includegraphics[width=0.38\textheight,angle=0]{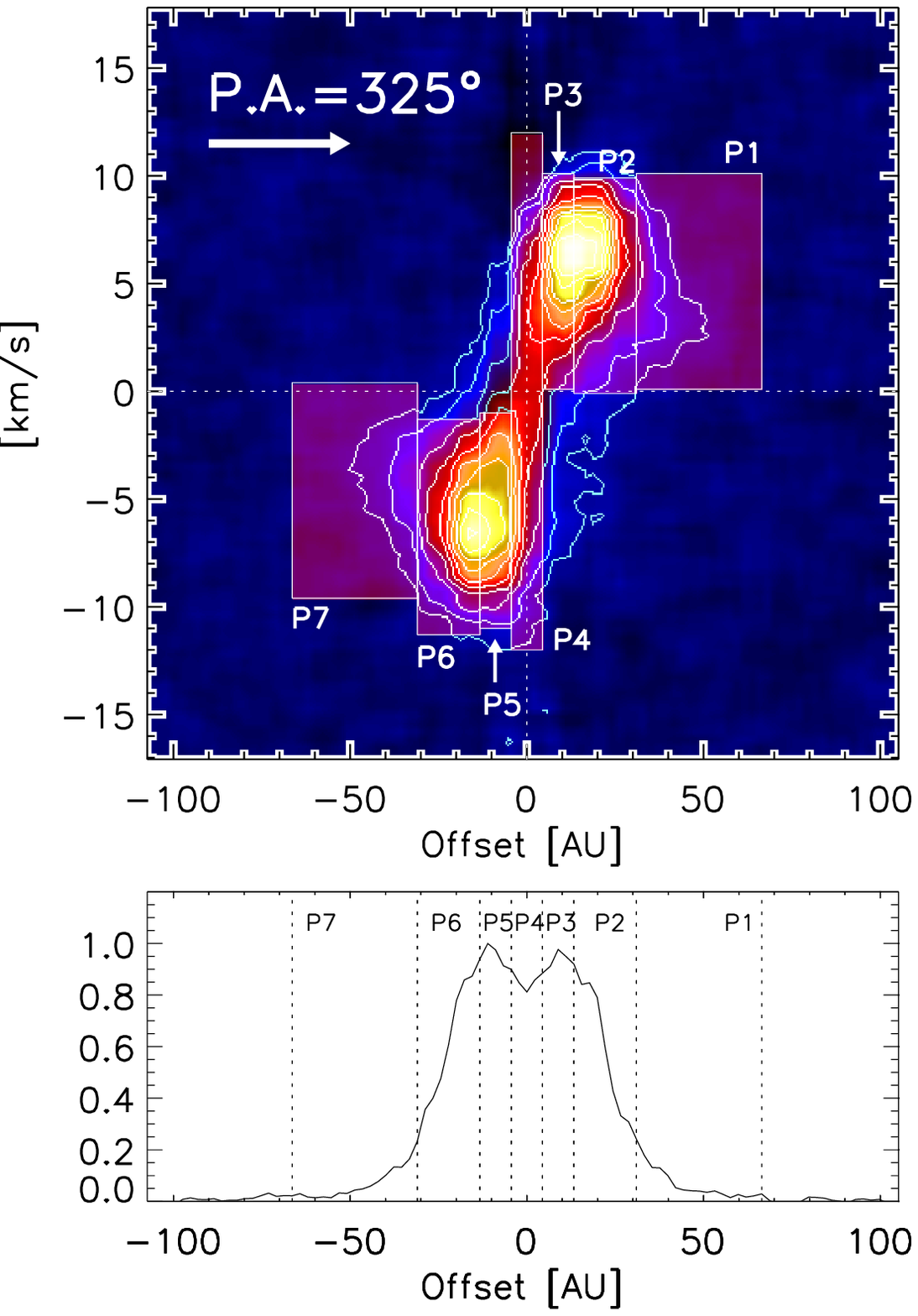} 
\caption{Top: Position-velocity diagrams of CO $v$=2-1 from
  HD~100546.  All vibrational transitions of CO $v=$2-1 in our
  observational coverage that do not overlap with the deep
  absorption lines of the terrestrial atmosphere are registered
  and added. The spectra were recorded with the slit
  P.A.=145\degr~(left) and 325\degr~(right). The spatial axis is
  in abscissa with the position angles to which the slit was
  aligned indicated by arrows. The line image obtained with slit
  P.A.=325\degr~is flipped horizontally so that the offset from
  the central star increases to the right in both of the line
  images. The continuum emission of the star was interpolated
  from both sides of the line emission and subtracted. The
  contours are drawn at every 10\% of the peak brightness,
  except the lowest one which is at 5\%.  The apertures of the
  spectral extraction are marked by rectangles.
  Faint emission is still visible at 50~AU away at low
  velocity. The sizes and the locations of the extraction
  apertures are summarized in Table~\ref{t1}.  Bottom: Spatial
  profile of the line emission, i.e., the position-velocity
  diagram shown in the upper panels are crushed along the
  velocity space.  \label{f1}}
\end{figure*}

\section{Data reduction and analysis}  

The spectral images were pre-processed for subtraction of
the dark-current images and flat-fielding, and coadded by
cries\_spec\_jitter recipe\footnote{CRIRES Pipeline User Manual
  VLT-MAN-ESO-19500-4406.} on the ESO gasgano
platform\footnote{http://www.eso.org/sci/data-processing/software/gasgano/.}.
Because all line images of $v=$2-1 look qualitatively similar,
those lines that were not blended with strong telluric
absorption lines, or with the other transitions of CO lines,
were registered and added to a single line image to
  increase the signal-to-noise ratio, and to check the spatial
extent of the gas emission (Fig.~\ref{f1}). Continuum emission
was interpolated from both sides of the line emission, and
subtracted. The gas emission of the disk is clearly resolved,
showing Keplerian rotation profiles consistently in both
line images obtained with slit P.A.$=$145\degr~and 325\degr~with
the low-velocity wings extending up to 50~AU.  The disk around
HD~100546 is known to be brighter at the southwestern side than
the northeastern side in the scattered light \citep{ard07}. The
asymmetry of the surface brightness has been attributed to the
preferential forward scattering of the dust grains in the
disk. The rotation of the disk detected in CO $v=$2-1 with the
southeastern part approaching toward us is consistent with this
disk inclination and the counter-clockwise winding of the spiral
arms \citep{gra05,ard07}.

One-dimensional spectra were extracted from the spectral images,
setting the extraction apertures at the incremental distances
from the central star (the positions and the sizes of the
  extraction apertures are summarized in Table~\ref{t1}). The
telluric absorption lines were removed by dividing by the
spectra of the spectroscopic standard star after correcting the
small mismatches in the wavelengths, the airmass, and the
spectral resolutions. The wavelength calibration was performed
to match the telluric absorption lines to the atmospheric
transmission model calculated by ATRAN \citep{lor92}.

\begin{figure*}
  \includegraphics[width=0.375\textheight,angle=0]{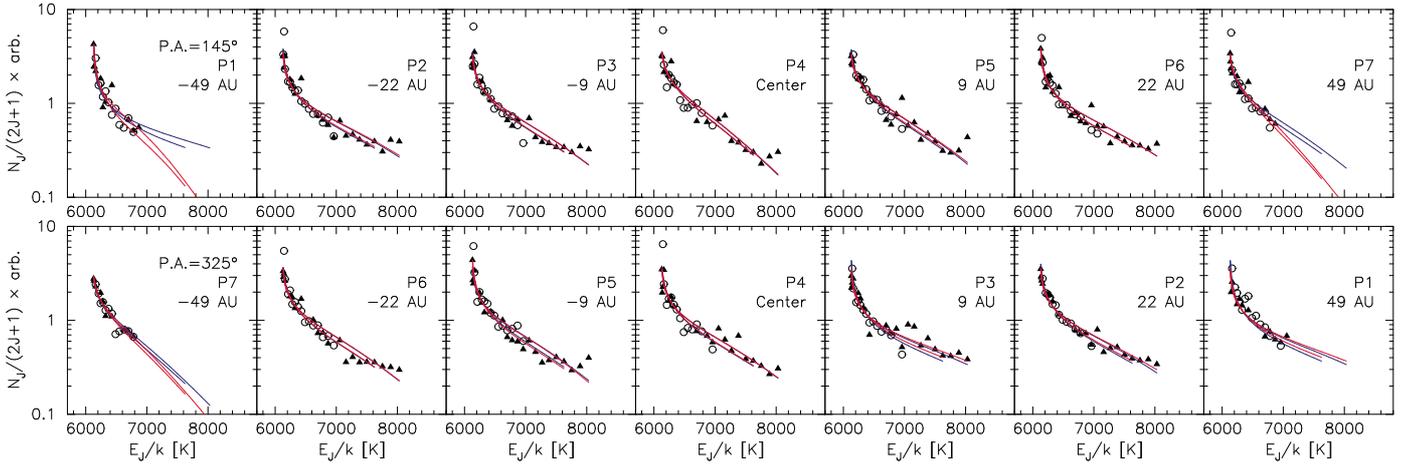}
\caption{Population diagrams of CO $v=$2-1 based on the spectra
  extracted at the incremental distances from the central
  star. The upper and the lower panels are from the spectra
  recorded with the slit P.A.=145\degr~and 325\degr. Open
  circles and filled triangles are for $R$- and $P$-branches,
  respectively. The column densities divided by the statistical
  weights are vertically shifted by arbitrarily amount for the
  presentation. The uncertainty of the individual column
    density is on the order of the  dispersions of the
    data points. The red lines are the best-fit LTE model
  population diagrams assuming Gaussian line width of
  6~km~s$^{-1}$. The blue lines are the same but the column
  density of CO is fixed to $N_{\rm
    CO}=10^{20}$~cm$^{-2}$.\label{f2}}
\end{figure*}

Because the transitions to the ground level $v=$1-0 were affected by
the telluric absorption lines of the same transitions, $v=$2-1
lines were used to calculate the column densities of the
rotational levels from $J$=0 to $J=$26 in $v=$2 vibrational
  state to construct the population diagrams. The column
  density at each $J$ level $ N_J$ was calculated for the
  optically thin limit $N_J \, A_{ul} \, h \nu \, \omega = W_\nu
  f_\nu $, where $\omega$ is the solid angle of the emitting
  area, and $f_\nu$ is the flux density of the continuum
  emission to convert the equivalent width $W_\nu$ to the line
  flux.  We were unable to perform the absolute flux calibration of
  the spectra to measure $f_\nu$ properly because of
  significant systematic errors from the variable PSF, variable
  slit transmission, and the variable pixel sampling of the PSF
  combined together.  The factor $f_\nu/\omega$ is irrelevant
  here, though, as long as it changes only slowly with the
  wavelength, because we are not primarily interested in the
  absolute column densities, but the relative distribution of
  $N_J$ to measure the excitation temperature of CO.  The line
  flux of the spectra obtained with adjacent grating settings
  were scaled so that the lines that were covered by both
  grating settings have the same line flux on average.  The
  column densities divided by the statistical weights are shown
  in Fig.~\ref{f2} after being shifted vertically by an arbitrarily
  chosen amount for the presentation.

Obviously, the gas is warm ($>$400~K) out to the
longest distance (49~AU) covered by the observations. In
  order to measure the excitation temperature of CO
  quantitatively, model population diagrams were calculated
  based on synthetic line emission spectra, and compared to the
  observations. The column densities in the upper and lower
  levels are given by $N_{(u,l)} = N_{v=(1,2)} \, g_{(u,l)} /
  Q_{(u,l)} \, e^{ -E_{(u,l)} / kT_{\rm ex}}$ in the LTE
  condition, where $N_{v=(1,2)}$ is the total CO column density
  at $v=$1 or 2 state, $g$ is the statistical weight, $Q$ is the
  partition function, $E_{(u,l)}$ is the energy of the upper or
  lower level with respect to the rotational ground level, and
  $T_{\rm ex}$ is the excitation temperature. The line source
  function of a two-level system was given by $S_\nu = N_{u}
  A_{ul} /(N_{l} B_{lu} - N_{u} B_{ul})$, and the line emission
  spectrum was therefore calculated as $I_\nu = S_\nu \, (1-
  e^{-\tau_\nu})$ with the line opacity $\tau_\nu = \frac{h \nu}
  { 4\pi } \, (N_{l} B_{lu} - N_{u} B_{ul})\, \phi(\nu)$.  The
  line profile $\phi(\nu)$ was assumed to be a Gaussian
  function.  We performed preliminary runs to fit the observed
  population diagrams by the models with limited sampling of
  $T_{\rm ex}$ and $N_{\rm CO}$, varying the line width of the
  Gaussian line profile from 2~km~s$^{-1}$ to 30~km~s$^{-1}$.
  The fitting result in terms of the absolute deviation ($
  \sum\limits_J|N_J^{\rm obs} - N^{\rm model}_J|$) became better
  up to 6~km~s$^{-1}$ in most of the spectra extracted from
  different disk locations. In several cases the fitting further
  improved up to 25~km~s$^{-1}$, but very slightly. For 
  simplicity, we fixed the line width to 6~km~s$^{-1}$, and used
  the same line profile for the spectra extracted at all 
  disk locations.  The equivalent widths of the emission lines
  were measured with the model spectra and converted to the
  column densities in the same way as the observed spectra.
  The energy levels, the line center wavelengths, and Einstein
  $A_{ul}$ coefficients were taken from \citet{goo94}. 
A more detailed discussion on the calculation can be found
elsewhere \citep{got11}. 

\begin{figure}
\includegraphics[width=0.32\textheight,angle=-90]{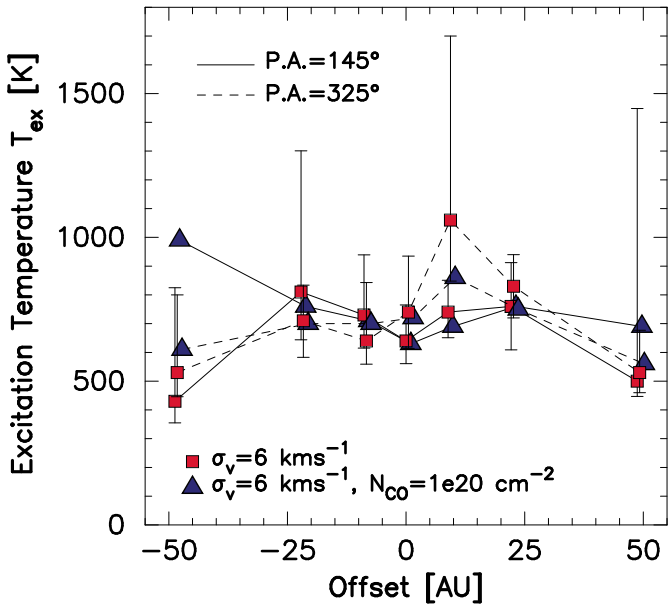} 
\caption{Radial profile of the rotational excitation
  temperature of CO $v=$2-1 in the disk atmosphere of HD~100546
  system. Data points are horizontally shifted slightly for
  clarity. The excitation temperatures are measured from the
  population diagrams presented in Figure~\ref{f2}. The red
  squares are for an LTE model with fixed line width of
  6~km~s$^{-1}$. The line emissions are optically thick,
  therefore the column density and the excitation temperature
  are highly degenerated. The error bars are given not by the
  fitting error in Figure~\ref{f2}, but by the total extent of
  the contour in temperature in the absolute deviation plots in
  Fig. \ref{fa1} and \ref{fa2}, where the absolute deviation
  becomes worse by 10~\% of the best fitting values. The blue
  triangles are from the fitting of the population diagrams, but
  with the column density of CO fixed to $N_{\rm
    CO}=10^{20}$~cm$^{-2}$.\label{f3}}
\end{figure}

The line emission is apparently optically thick with turnovers
seen around $E_J/k\sim$6500~K. The excitation temperature and
the column density are heavily degenerated in these cases. The
formal fitting error might underestimate the realistic
uncertainty in the temperature measured. We produced contour
maps of the absolute deviations between calculated and observed
population diagrams on a grid of temperatures (500--2500~K) and
column densities (10$^{19}$--10$^{22}$~cm$^{-2}$)
(Fig.~\ref{fa1}, \ref{fa2} in the online edition).  The total
extents of the contour in temperature where the absolute
deviation increases by 10\% of the best fitting value were taken
as the uncertainties of the excitation temperature of CO
(Fig.~\ref{f3}, Table~\ref{t1}).  We lack the measurements of
the equivalent widths of high-$J$ levels at 49 AU because of the
insufficient signal-to-noise ratio of the spectra
(Fig.~\ref{f2}).  The upper limit of the excitation temperature
is therefore less constrained at the radius than at other
locations, but the conclusion ($T_{\rm ex} >$400~K) is not
affected. Fitting was also performed with the column densities
fixed to $10^{20}$~cm$^{-2}$ to give an independent estimate of
how the result is affected by this degeneracy. The resulting
excitation temperatures do not differ by more than the extent of
the errors estimated above except at $-$49~AU with
P.A.=145\degr.

\section{Discussion}

The rotational excitation temperature of CO is between 400~K and
1100~K at all locations on the disk where the spectra were
extracted. There might be a hint of the cooling of the gas down
to 400--600~K with the distance from the central star, but this
is not unambiguous. The measurements are consistent in the data
with P.A.=145\degr~and 325\degr, except at the edge of the disk
at $\pm$49~AU, and the inner disk at $+$9~AU from the star. The
latter might be picking up the emission from the hot inner rim
exclusively. Although the observation was made with high angular
resolution, the line emission from different radii of the disk
overlap up to 22~AU, because the emission from the inner rim is
by far stronger than that from the disk behind (the spatial
profiles of the line emission are shown in the lower panels in
Fig.~\ref{f1}). However, as is seen in the locations of the
extraction apertures overlaid with the line images
(Fig.~\ref{f1}), the spatial resolution is sufficiently high
that little blending with the emission from the inner disk is
expected at 50~AU and beyond. The excitation temperature is
still hotter than 400~K at 49~AU, where the blackbody
equilibrium temperature with the stellar radiation [$\sigma\,
T_{\rm eq}^4=L_\ast/(16\,\pi\, d^2)$; $\sigma$, $L_\ast$, $d$
are the Stefan-Boltzmann constant, the luminosity of the star,
and the distance from the star to the particle] drops as low as
90~K. Although the presence of this warm gas is exactly what the
disk model predicts \citep{kam04,jon04}, this is the first time
that the temperature of the disk atmosphere is directly measured
beyond the disk inner rim by a high angular resolution
observation that spatially resolved the disk. The range of the
excitation temperature nicely overlaps with 300--800~K of the
pure rotational lines of CO at high-$J$ (14$\le J \le$30)
measured in the recent spectroscopy of the same object by {\sc
  Herschel}/PACS \citep{stu10}. \citet{thi11} also detected
CH$^+$ $J$=5-4, 6-5, 3-2 by the same instrument, and derived a
rotational excitation temperature 323$^{+2320}_{-150}$~K.  It is
inferred that these observations are likely looking at the gas
in a similar location of the disk as the present study. The
model presented by \citet{thi11}, which takes into account the
specific disk geometry of HD 100546 constrained by the spectral
energy distribution \citep{ben10}, also shows a warm layer of
the gas at 50 AU of the disk.

A mechanism that likely vibrationally excites CO to $v=$2 level
is the ultraviolet pumping \citep{kro80}, because the
vibrational temperature \citep[$T_{\rm vib}=6600\pm 700$~K;
][]{pla10} is much hotter than the rotational excitation
temperature \citep{bri09}. The rotational levels are much faster
to thermalize through the collisions with the molecular/atomic
hydrogen and therefore should represent the ambient gas
temperature better.  It is in principle possible to account for
the extended line emission not by the emission arising on the
site, but by the scattered line emission, where CO is excited to
$v=$2 level in the vicinity of the central star, and the line
emission emerging from the hot inner disk is simply reflected
into the line of sights to observers by the dust grains at 50
AU. We discarded the scattered line emission, because the line
images in Fig.~\ref{f1} clearly show the Keplerian profiles,
which indicate that the line center velocity at 50~AU is closer
to the system velocity than that of the gas orbiting closer to
the central star.

The observed excitation temperature agrees with that of
\citet{jon07}, who calculated the chemistry and the temperature
structure of a disk around a Herbig~Ae star without the a priori
assumption that the gas is physically coupled to the dust. The
calculations were performed for the disk mass from $10^{-1}$ to
$10^{-4} M_\odot$ to simulate the dissipation of a disk, and
with and without the dust settling, i.e., the mass ratio of the
dust and the gas in the upper layer gradually reduced from 0.01
to 10$^{-5}$. The gas temperature in the upper layer is clearly
higher in cases without dust settling, because the gas is
primarily heated by the photoelectric effect of the dust grains.
The gas is hotter than 300~K in the disk atmosphere at the
distance of 50~AU from the disk surface to the significant depth
toward the mid-plane for all the disk models calculated without
dust settling, which is comfortably matched with the excitation
temperature we found; while the gas temperature with dust
settling is everywhere less than 100~K. That the disk atmosphere
is well mixed with the dust grains is also consistent with the
classification of the source as Ia by \citet{meu01} with
spatially resolved \citep{hab06}, rich emission features of PAHs
\citep{ack10}, and the thermal continuum emission extended up to
1\farcs4, or 145~AU, in the mid-infrared \citep{mul11}.

\begin{acknowledgements}
  We appreciate the constructive criticisms of the anonymous
  referee that improved the manuscript.  We thank all the staff
  and crew of the VLT for their valuable assistance in obtaining
  the data.  We appreciate the hospitality of the Chilean
  community that made the research presented here possible.
\end{acknowledgements}

\appendix 

\section{Measurement of rotational excitation temperature}

Here we provide the contour maps of the absolute deviations
between calculated and observed population diagrams 
that we
  used to estimate the uncertainty of the excitation temperature
  of CO.  
  The extents of the contour in the temperature where the
  absolute deviation increases by 10\% of the best fitting value
  were taken as the error of $T_{\rm ex}$.

\onlfig{1}{
\begin{figure*}
\includegraphics[width=0.3\textheight,angle=0]{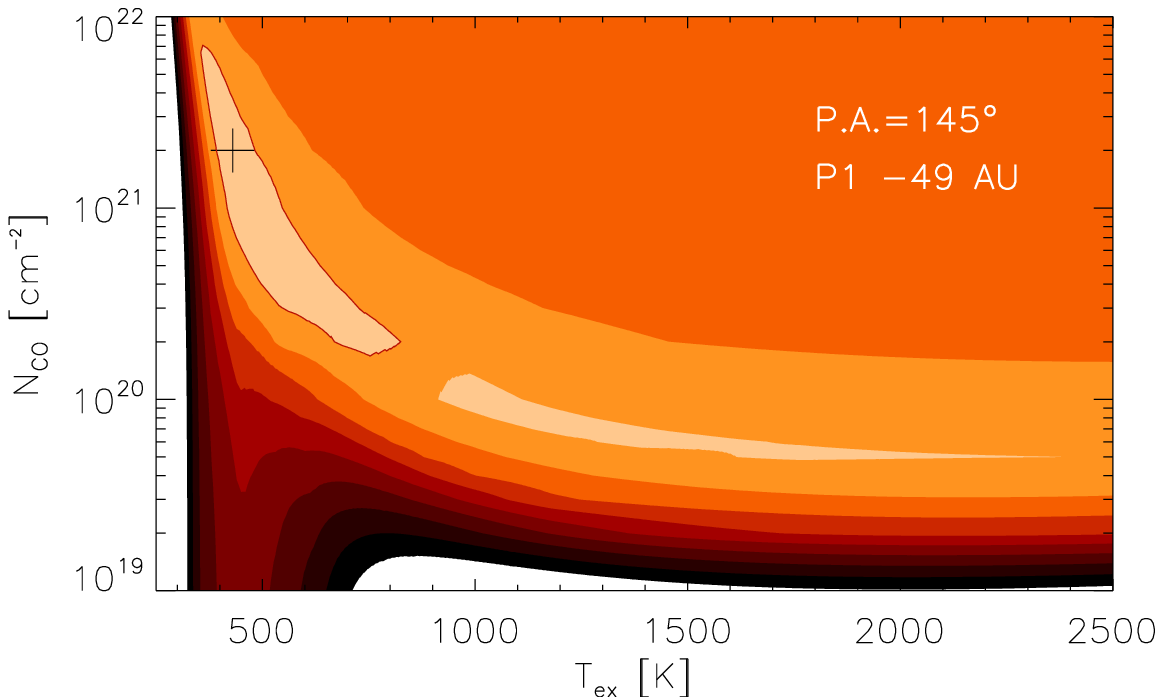}
\includegraphics[width=0.3\textheight,angle=0]{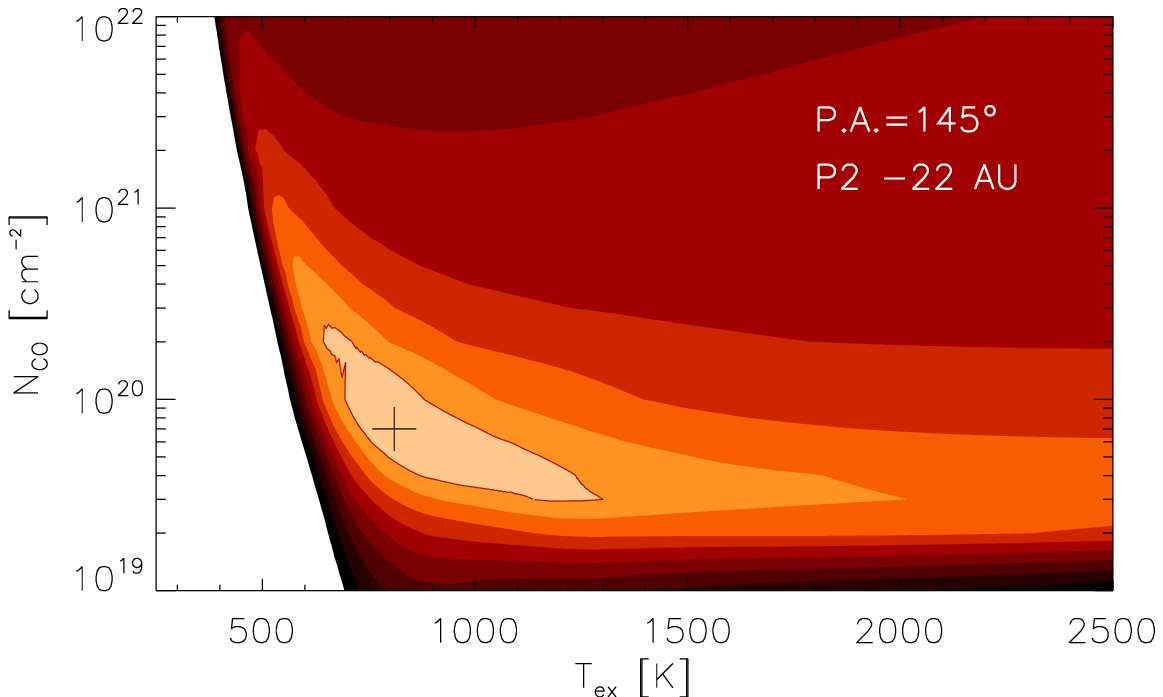}
\includegraphics[width=0.3\textheight,angle=0]{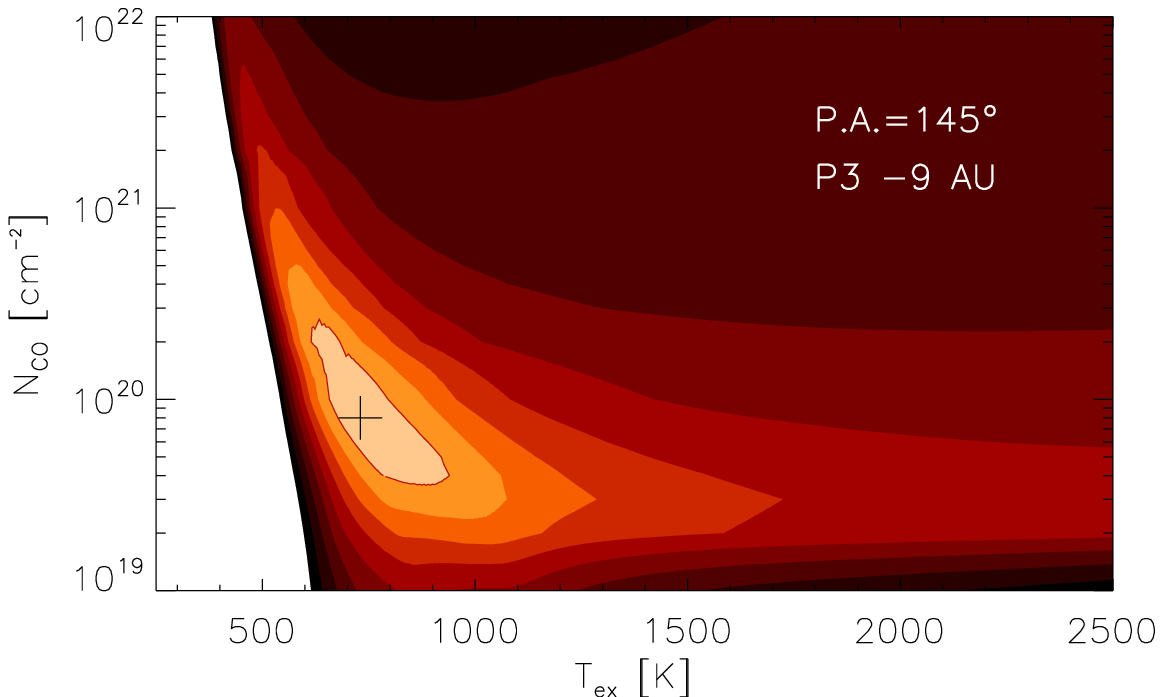}
\includegraphics[width=0.3\textheight,angle=0]{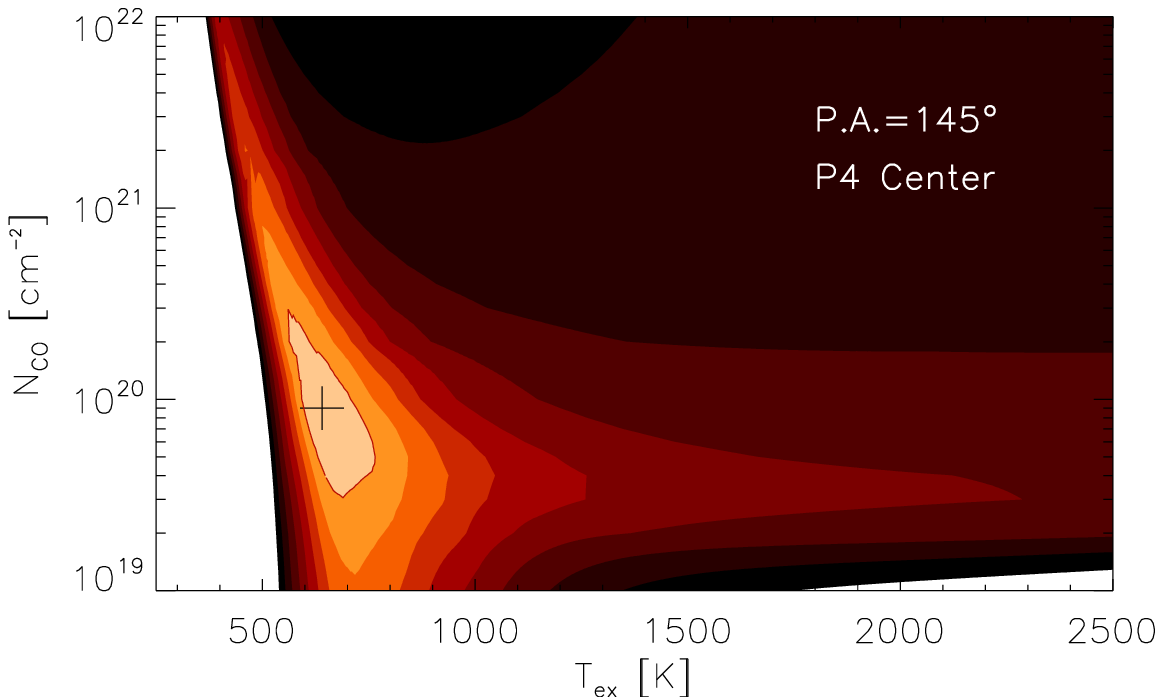}
\includegraphics[width=0.3\textheight,angle=0]{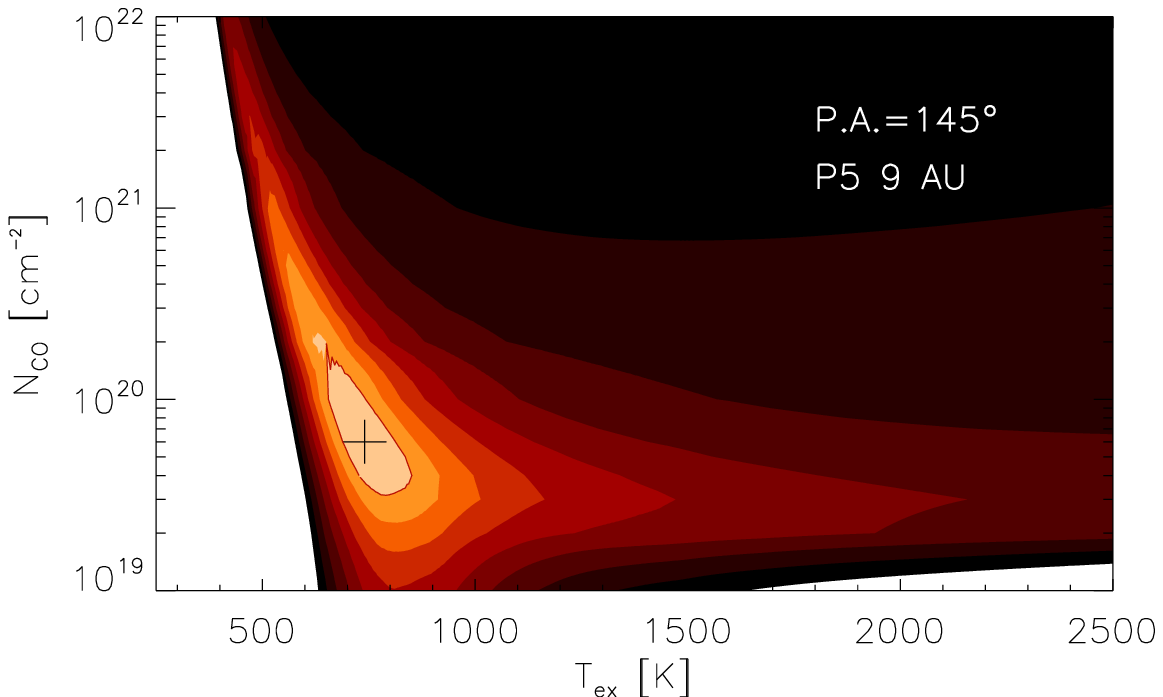}
\includegraphics[width=0.3\textheight,angle=0]{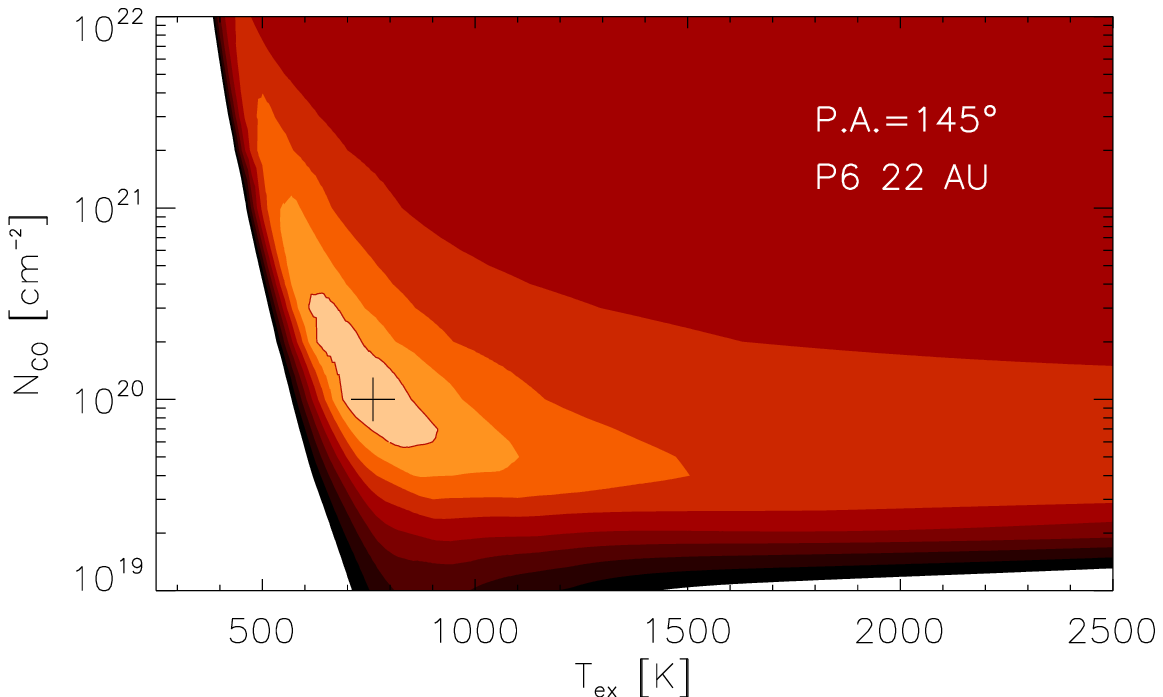}
\includegraphics[width=0.3\textheight,angle=0]{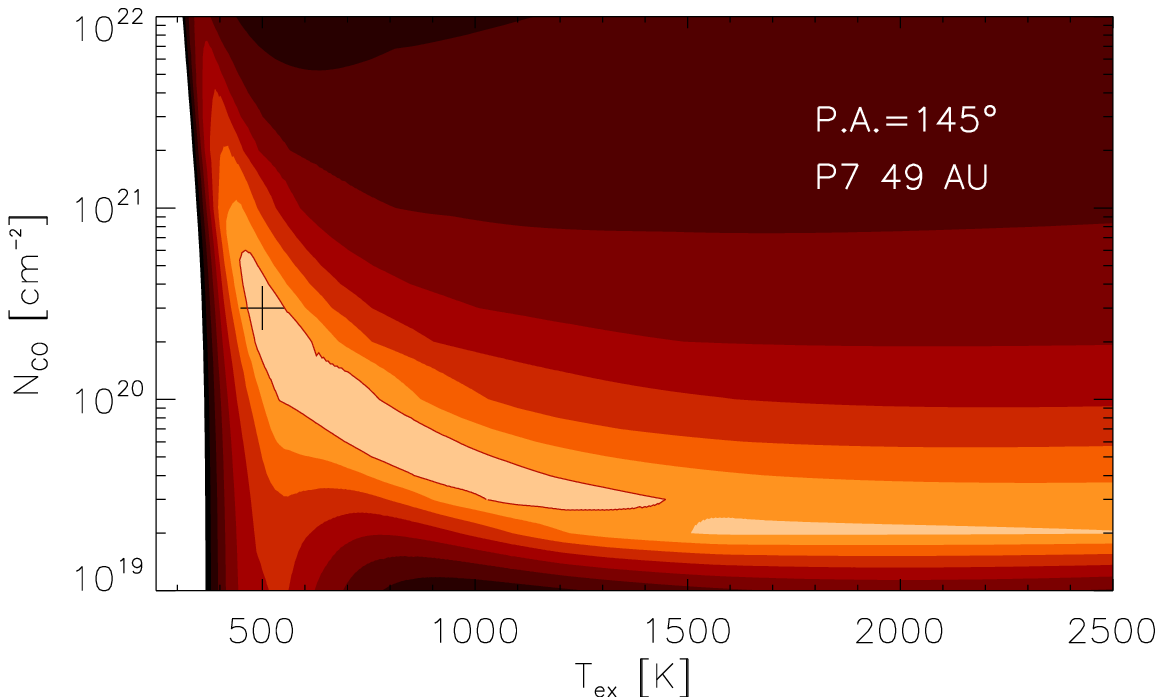}
\caption{Contour maps of the absolute deviations between the
  calculated and observed population diagrams. The observed
  population diagrams are based on the data obtained with the
  slit P.A.=145\degr.  $T_{\rm ex}$ and $N_{\rm CO}$ that give
  the minimum absolute deviations are marked by crosses. The
  contours are drawn at every 10\% increase of the absolute
  deviation from its minimum value. \label{fa1}}
\end{figure*}
}

\onlfig{2}{
\begin{figure*}
\includegraphics[width=0.3\textheight,angle=0]{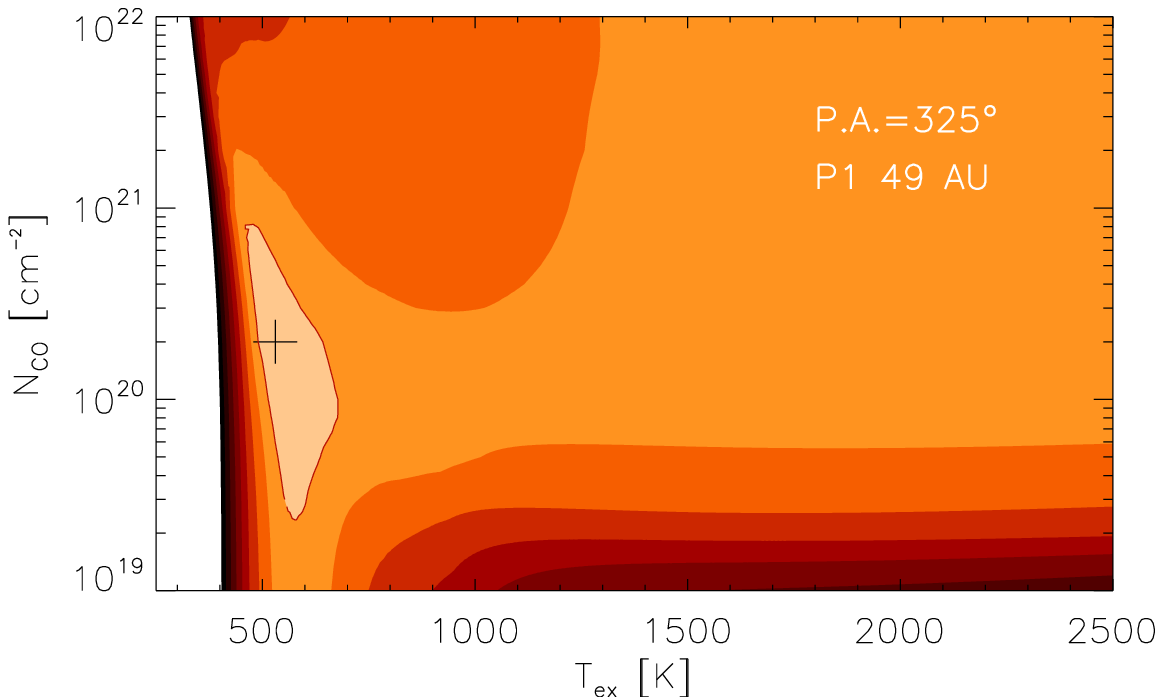}
\includegraphics[width=0.3\textheight,angle=0]{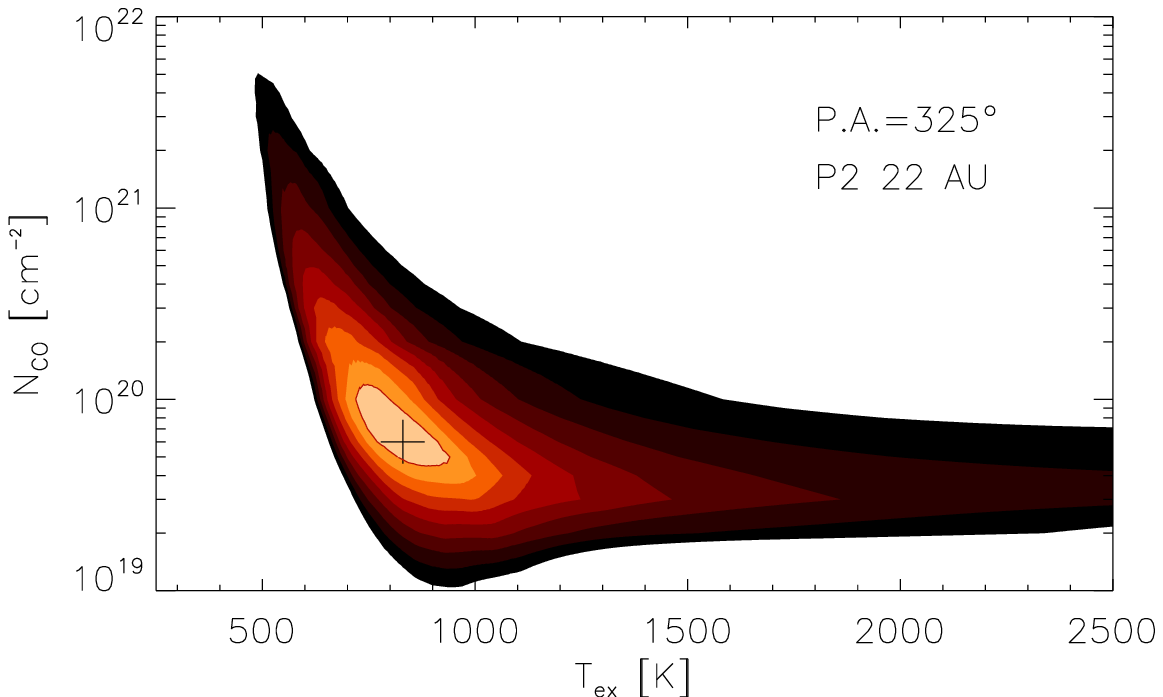}
\includegraphics[width=0.3\textheight,angle=0]{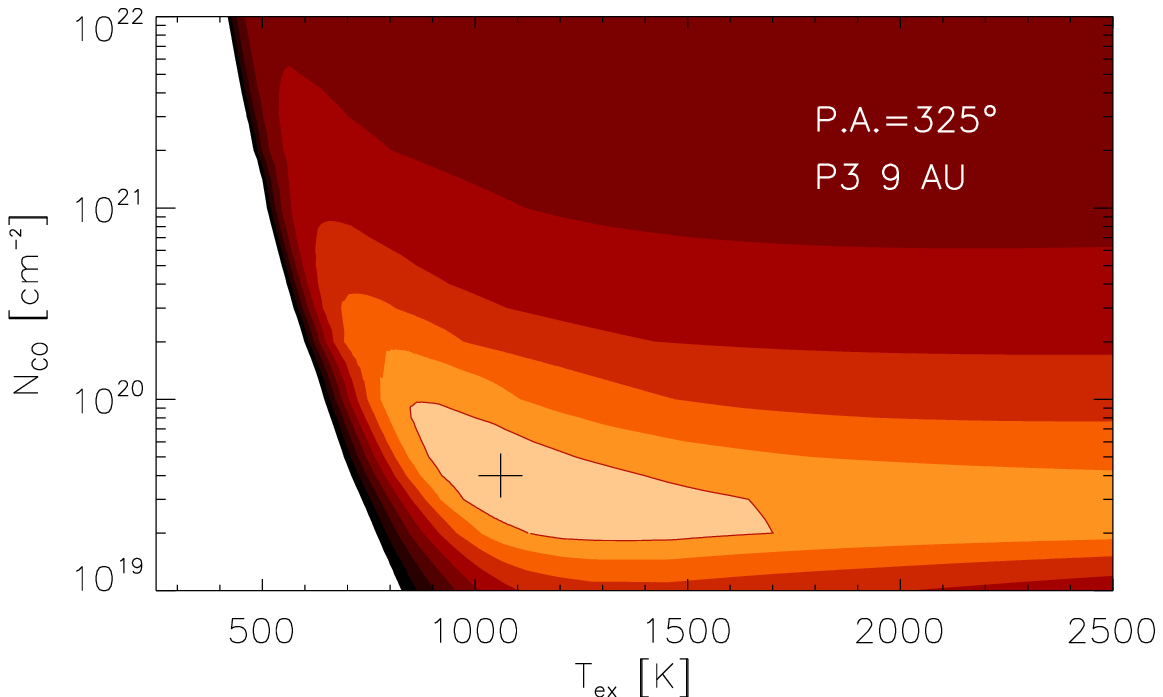}
\includegraphics[width=0.3\textheight,angle=0]{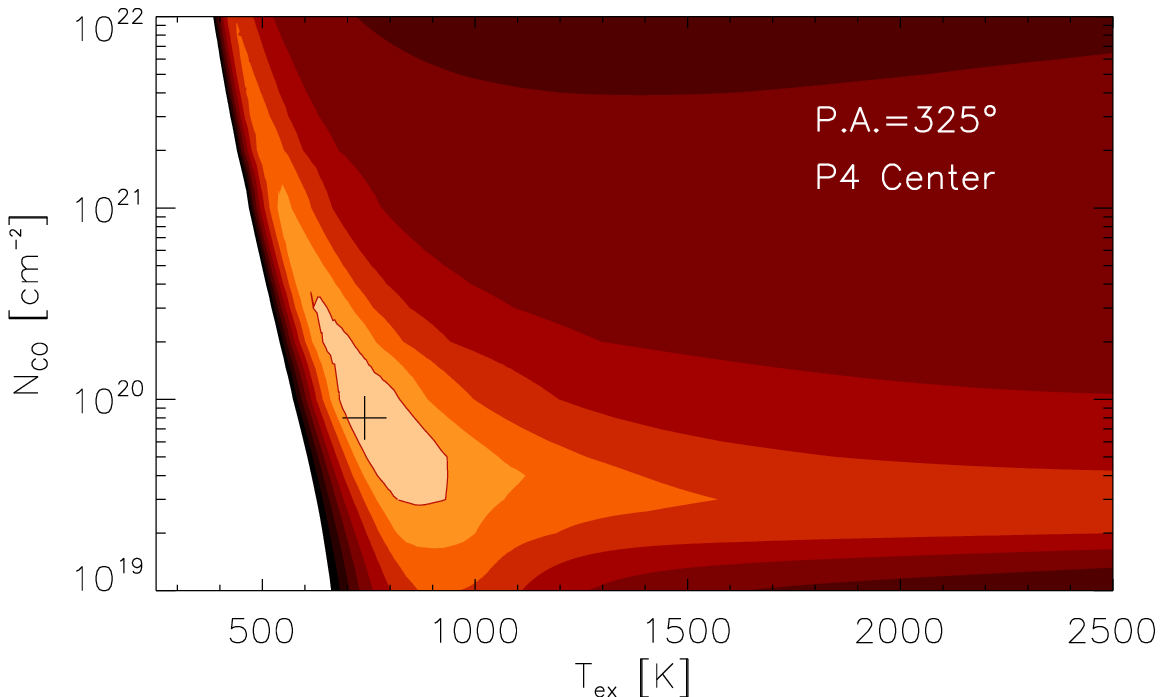}
\includegraphics[width=0.3\textheight,angle=0]{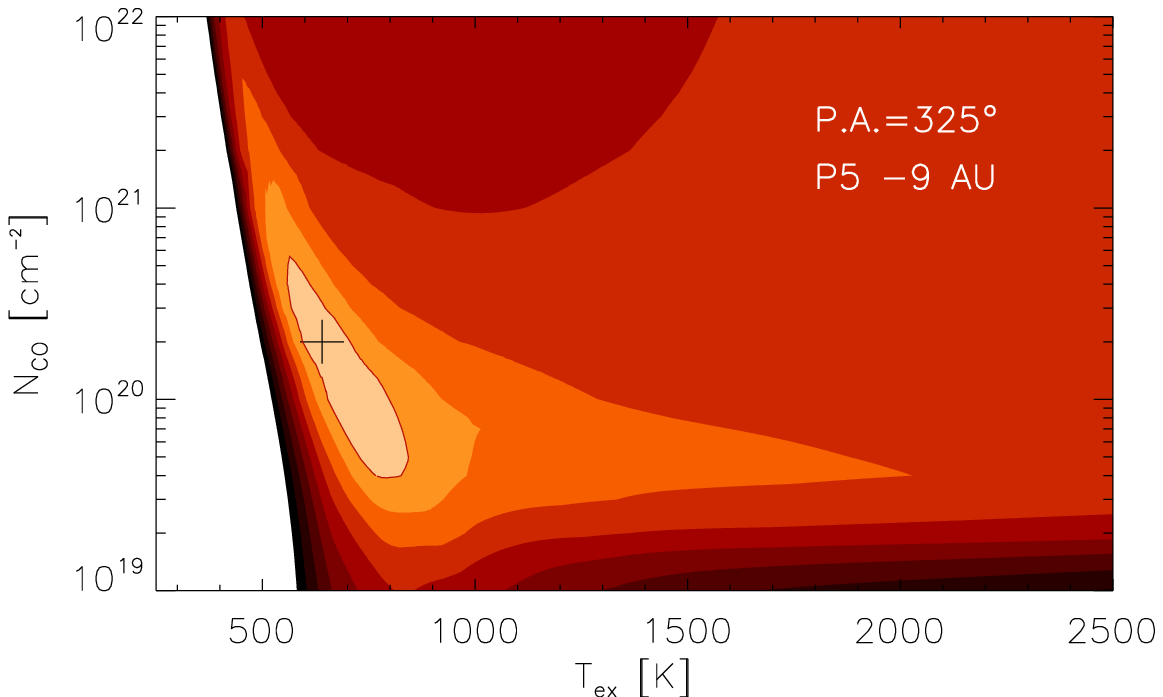}
\includegraphics[width=0.3\textheight,angle=0]{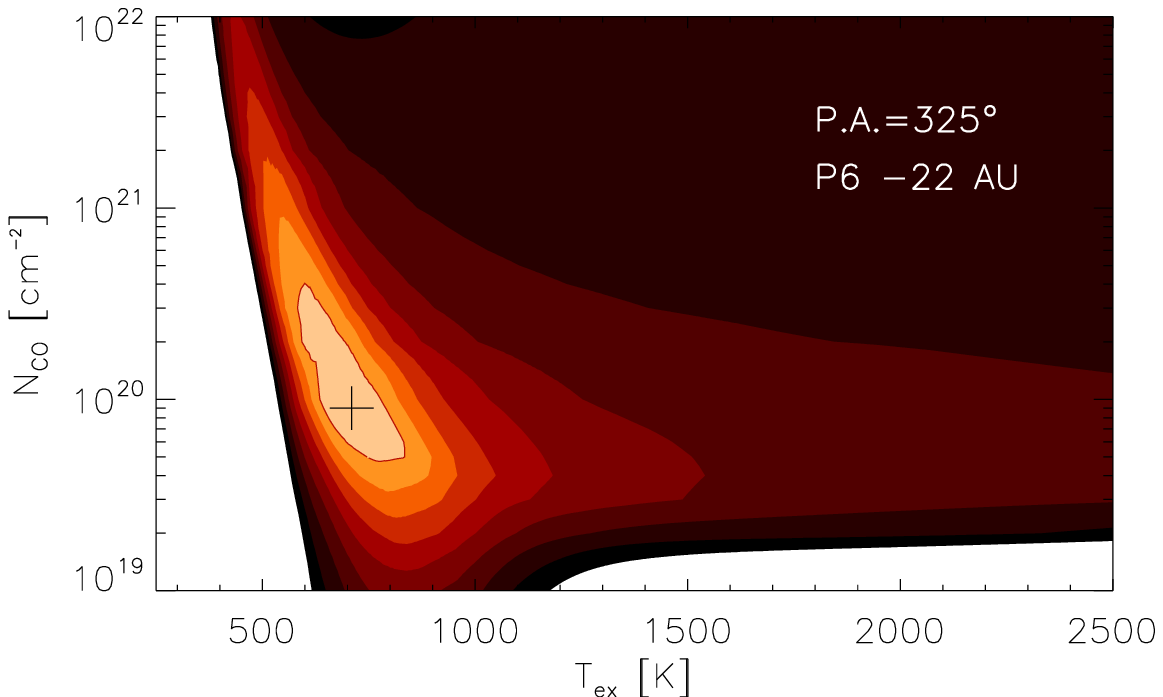}
\includegraphics[width=0.3\textheight,angle=0]{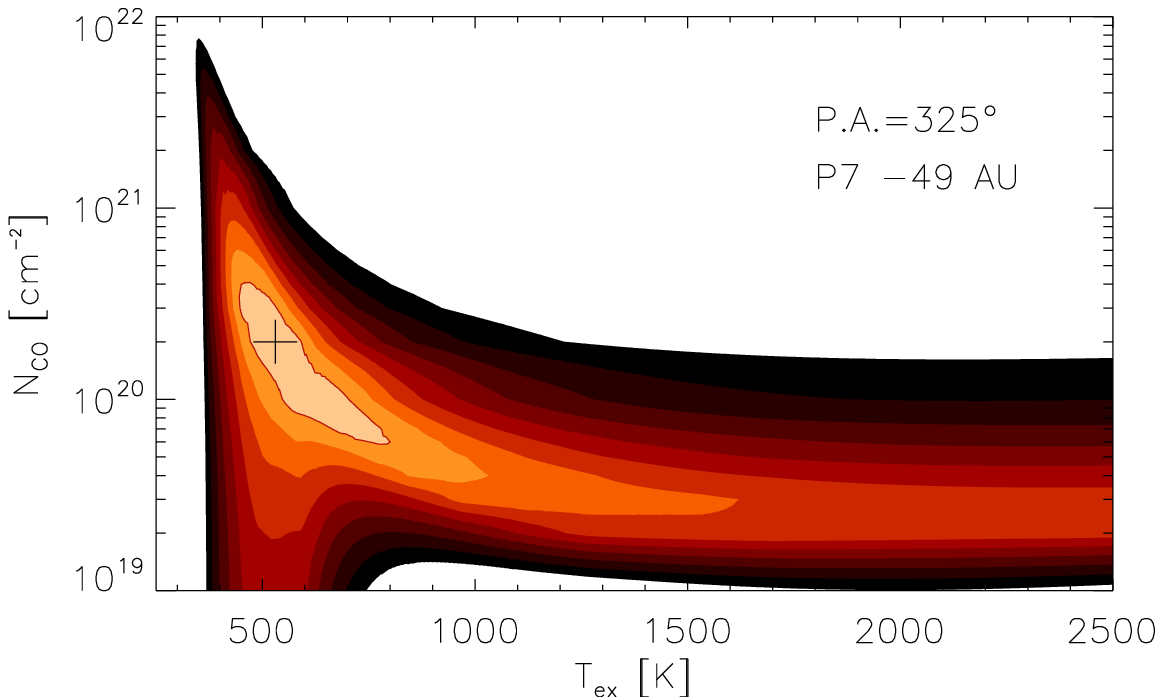}
\caption{Same as Fig.~\ref{fa1}, but for the data obtained 
with the slit P.A.=325\degr.\label{fa2}}
\end{figure*}
}

\end{document}

%% file: t1.tex
\begin{table*}
\caption{Extraction apertures and excitation temperatures.}
\label{t1}
\centering
\begin{tabular}{lc rrrrrrr}
\hline \hline

                    P.A.=145\degr\dots                &             & P1            & P2           & P3           & P4         & P5          & P6           & P7     \\
\hline
                    Aperture location\tablefoottext{a}& [pix]       & $-$5.5 (4.0)  &$-$2.5 (2.0) & $-$1.0 (1.0) & 0.0 (1.0) & 1.0 (1.0)   & 2.5 (2.0)    & 5.5 (4.0)      \\
                    Aperture location\tablefoottext{b}& [AU]        &$-$48.7 (35.4) &$-$22.1(17.7) & $-$8.9 (8.9) & 0.0 (8.9) &8.9 (8.9) &22.1 (17.7) & 48.7 (35.4)  \\
                    Line center\tablefootmark{c}   &[km~s$^{-1}$]&$-$4.9($\pm$5.0)   & $-$6.3($\pm$5.0)& $-$6.0($\pm$5.0) & 0.0($\pm$12.0) & 5.5($\pm$5.0)   & 6.0($\pm$5.0)    & 5.1($\pm$5.0)        \\

                                                      &             &        &        &        &      &       &        &        \\
                    $T_{\rm ex}$\tablefootmark{d}     & [K]         &430$^{+395}_{-75}$&810$^{+491}_{-166}$&730$^{+209}_{-115}$&640$^{+125}_{-79}$&740$^{+111}_{-89}$&760$^{+152}_{-151}$&500$^{+948}_{-53}$\\

\hline \hline

                    P.A.=325\degr\dots                &             & P7            & P6           & P5           & P4         & P3          & P2           & P1     \\
\hline
                    Aperture location\tablefoottext{a}& [pix]       & $-$5.5 (4.0)  &$-$2.5 (2.0) & $-$1.0 (1.0) & 0.0 (1.0) & 1.0 (1.0)   & 2.5 (2.0)    & 5.5 (4.0)      \\
                    Aperture location\tablefoottext{b}& [AU]        &$-$48.7 (35.4) &$-$22.1(17.7) & $-$8.9 (8.9) & 0.0 (8.9) & 8.9 (8.9) &22.1 (17.7) & 48.7 (35.4)  \\
                    Line center\tablefootmark{c}   &[km~s$^{-1}$]&$-$4.6($\pm$5.0)     & $-$6.3($\pm$5.0)   & $-$6.0($\pm$5.0)   & 0.0($\pm$12.0)   & 5.1($\pm$5.0)     & 5.9($\pm$5.0)      & 5.1($\pm$5.0)        \\
                                                      &             &        &        &        &      &       &        &        \\
                    $T_{\rm ex}$\tablefootmark{d}     & [K]         &530$^{+270}_{-84}$&710$^{+124}_{-127}$&640$^{+203}_{-81}$&740$^{+195}_{-126}$&1060$^{+640}_{-212}$&830$^{+110}_{-110}$&530$^{+147}_{-70}$\\

\hline

\end{tabular}
\tablefoottext{a}{Location of the aperture center at the spectral extraction. The positions are given  with respect to that of the central star in the unit of detector pixel. The numbers in parenthesis are 
  the sizes of the apertures.}
\tablefoottext{b}{Positions and sizes of the apertures in pixel unit are converted to the physical scale assuming the distance to HD~100546 being $d=$103~pc, and the pixel scale of CRIRES~86~mas.}
\tablefoottext{c}{Line center and full interval of the velocity range that the intensity of the line emission is integrated to calculate the line flux. Velocity offset is given with respect to the line center at the aperture location P4.} 
\tablefoottext{d}{The measurements are from the fittings of the population diagrams with LTE models presented in the red lines  in Figure~\ref{f2}.}
\end{table*}